# Dynamic evolution of internal stress, grain growth, and crystallographic texture in arc-evaporated AlTiN thin films using in-situ synchrotron x-ray diffraction

Sanjay Nayak[1,*], Tun-Wei Hsu[1], Robert Boyd[1], Jens Gibmeier[2], Norbert Schell[3], Jens Birch[1], Lina Rogström[1], and Magnus Odén[1]

[1]Department of Physics, Chemistry and Biology (IFM), Linköping University, SE-581 83 Linköping, Sweden

[2]Institute for Applied Materials, Materials Science and Engineering (IAM-WK), Karlsruhe Institute of Technology, 76131 Karlsruhe, Germany

[3]Helmholtz-Zentrum Hereon, Institute of Materials Physics, Max-Planck-Str. 1, 21502 Geesthacht, Germany

## Abstract

Understanding the nucleation and growth of polycrystalline thin films is a long-standing goal. Polycrystalline films have many grains with different orientations that affect thin-film properties. Numerous studies have been done to determine these grain size and their preferred crystallographic orientation as well as stress in films. However most past studies have either employed an ex-situ methodology or only monitor the development of macroscopic stress in real-time. There has never been any research done on the simultaneous determination of crystallographic texture, grain size, and microscopic stress in polycrystalline thin films. In this study, we simultaneously monitored the generation and temporal evolution of texture, grain size, and internal stress in cathodic arc evaporated $Al_{0.50}Ti_{0.50}N$ thin films using a bespoke deposition apparatus designed for use with 2-dimensional synchrotron x-ray diffraction technique. The influence of the substrate temperature is investigated in terms of the emergence and development of texture, grain size and stress evolution. A dynamic evolution of the crystallographic texture is observed as the overall film thickness varies. We clearly resolved two regime of films growth based on stress evolution. Beyond a threshold grain size (~ 14 nm), the stress scales inversely to the average grain sizes, and as the film thickness increases, immediate compressive stress relaxation was seen. An extensive ex-situ evaluation of thin films using electron microscopies and electron diffraction was performed to support the in-situ x-ray diffraction results.

*Corresponding author.
Email address: sanjay.nayak@liu.se, sanjaynayak.physu@gmail.com (SN)

## I. Introduction

Transition metal nitrides (TMNs), especially AlN-alloyed TiN ($Al_xTi_{1-x}N$)-based thin films, show extraordinary physical, mechanical, and tribological properties, e.g., high mechanical hardness, high melting temperature, thermodynamic stability at elevated temperature, and oxidation resistance [1,2]. Owing to these characteristics, thin films of $Al_xTi_{1-x}N$ are preferred as protective films that extend the life of cutting tools and inhibit corrosion. In practice, polycrystalline thin films of $Al_xTi_{1-x}N$ are deposited on top of a rigid substrate (e.g., cutting tools) by physical vapour deposition (PVD), such as magnetron sputtering or cathodic arc evaporation, or chemical vapour deposition (CVD) techniques. Typical PVD deposited films possess nanocolumnar and/or nanocrystalline morphologies [3] with preferred orientation of certain crystallographic planes along the growth direction (known as crystallographic texture). As the PVD deposition involves non-equilibrium processes and the formation of high-energy ions, as-deposited films are also often in a state of residual stress. These inherent stresses (usually compressive stresses) in PVD coated thin films play a vital role in the overall performance and efficiency of the films. During mechanical operations, the stress due to external mechanical loads adds to the pre-existing stress. A high pre-existing stress may lead to the complete failure of the coating, including its delamination from the substrate. At the same time the smaller values of residual stress and low values of mechanical hardness correlate with each other[4,5]. Thus, optimizing the stress in thin films is important and it is required for industry to realize their full potential. In this regard understanding the mechanisms of thin film nucleation and growth has been an important aspect of research for the community, since these factors are essential in determining the properties of films and their practical significance.

It has already been established that many process parameters in PVD can be used to tune the film's stress level. For example, application of a negative substrate bias voltage ($V_s$) leads to the formation of finer grains and an increase in residual stress [6–9]. At the same time, higher $V_s$ also helps in the deposition of dense or compact films and leads to high mechanical hardness. The partial pressure of $N_2$ in the total growth pressure (Ar+$N_2$) influences the composition, roughness, grain size, and microstructure of the films. The non-stoichiometric films, resulting from different Ar+N2 combinations, show poor mechanical hardness [10–13]. Machet et al. [14] examined the effect of substrate temperature (varying from 100 °C to 450 °C) on the elemental composition, surface morphology, and mechanical hardness properties of magnetron sputtered TiN films. They found that for a given $N_2$ partial pressure, the development of crystallographic texture is greatly influenced by the substrate temperature. At low growth temperatures (below 300 °C), 111 crystallographic texture of the face-centered-cubic (FCC) TiN phase forms in the growth direction (GD), while at higher temperature a 200 and/or 220 texture of FCC-TiN are developed along GD. Machet et al. characterized the mechanical properties of the films and claimed that TiN films with a crystallographic texture of 111 possess a hardness of 20 GPa, but exhibit poorer wear resistance than films with a crystallographic texture of 200, which have a hardness of 25 GPa along with a good wear resistance [15]. A similar observation was made by Wang et al. [16] in (Ti, Al)N films deposited with cathode discharge deposition process. Kalss et al. investigated the interdependence of stress and crystallographic texture in (Al, Ti)N films deposited with cathodic arc evaporations [17] and argued that the formation of high index crystallographic plane's texture (e.g., 113 and 115) leads to residual stress relaxation and reduction in film density, contrary to Beckers et al. [18] and Greene et al [19].

Most of the above-mentioned reports determined residual stress (or simply stress), crystallographic texture, and grain size after the film's deposition and after cooling down from growth to room temperature. Little is known about the real-time formation and evolution of stress and microstructure of $Al_xTi_{1-x}N$ thin films, particularly when deposited with the industry-preferred arc evaporation technique. Recently, by utilizing a home designed cathodic arc evaporation chamber [20], adapted for in-situ synchrotron radiation studies, we recorded the evolution of stress/strain and microstructure of polycrystalline $Al_xTi_{1-x}N$ (x = 0, 0.25, 0.5, and 0.67) thin films under different substrate bias voltage ($V_s$) [21]. In this report, we examine the formation and evolution of biaxial stress, crystallographic texture, and grain size in $Al_{0.50}Ti_{0.50}N$ films deposited by varying substrate temperatures, while maintaining a constant substrate electrical bias of -40 V.

## II. Experimental Details

To deposit $Al_{0.50}Ti_{0.50}N$ films on the 100 surface of a Si substrate, a reactive cathodic arc evaporation technique was used. A 63 mm diameter $Al_{0.50}Ti_{0.50}$ metallic cathode (FK Grade, PLANSEE-Germany) was used at a fixed arc current of 75 Amperes. The angle formed by the surface normal of the cathode and the normal of the substrate was 35°. A substrate rotation speed of 5 rotations per minute was maintained. The details of the experimental setup were previously discussed [20,21]. The pure $N_2$ (99.9995 %) pressure was kept constant at 40 mTorr during depositions. The substrate was heated using an 808 nm infrared laser diode-based heating unit. In this study, four different substrate temperatures ($T_s$) were used for thin film deposition: 250, 450, 600, and 750 °C. A DC power supply was used to electrically bias the substrate with a fixed voltage of -40 V.

The deposition system was placed at the P07 beamline (DESY, Germany) for in situ x-ray diffraction (XRD) during growth. The deposition system was aligned such that the x-ray beam travelled through the growing coating, parallel to the substrate surface. The x-ray energy was 73.79 keV and the beam size was defined by slits to 100×600 µm². A 2D flat panel detector with a 2048×2048-pixel resolution was used to record the diffractograms (PerkinElmer XRD 1622). Each exposure was 0.2 seconds, and ten exposures were superimposed to create the final diffractogram. The diffractogram of NIST standard $LaB_6$ powder was used to measure the distance between the substrate and the detector. The detector was placed to capture one quadrant of the diffraction pattern to maximize resolution. The 2D diffractograms were converted into 1D-line profiles using pyFAI [22], a Python-based program that integrates the intensities using 5° wide azimuthal bins ($\Delta\varphi$). To obtain the peak positions and peak widths of the diffraction peaks, the intensity vs. 2θ data was fitted with a pseudo-Voigt function. The biaxial stress was determined by the well-known $sin^2\psi$ method. The unstrained interplanar spacing ($d_{hkl}^0$) was determined from the $d_{hkl}$ vs. $sin^2\psi$ plot and is corresponding to $d_{hkl}$ value at the corresponding invariant tilt angle $\psi^*$, i.e., $d_{hkl}^0 = d_{hkl}(\psi = \psi^*)$. The $\psi^*$ was determined by the relation: $sin^2\psi^* = \frac{2\nu_{hkl}}{1+\nu_{hkl}}$, here $\nu_{hkl}$ is the Poisson ratio for the hkl orientation of the films. The 111-diffraction peak was used to determine the biaxial stress. High temperature elastic constants of FCC- $Al_{0.50}Ti_{0.50}N$ from the work of Shulumba et al. [23] were used to estimate strain and stress and are tabulated in Table 1.

Table 1: Temperature dependent elastic coefficients of FCC- $Al_{0.50}Ti_{0.50}N$ used in determination of biaxial stress in thin films. The values are obtained from the work of Shulumba et al. [23].

| $T_s$ (°C) | $C_{11}$ (GPa) | $C_{12}$ (GPa) | $C_{44}$ (GPa) | $E_{111}$ (GPa) | $\nu_{111}$ | $\psi^*$ (°) |
|---|---|---|---|---|---|---|
| 250 | 410 | 134 | 178 | 422.9 | 0.188 | 32.5 |
| 450 | 394 | 130 | 171 | 406.67 | 0.189 | 32.5 |
| 600 | 382 | 127 | 166 | 394.9 | 0.189 | 32.6 |
| 750 | 369 | 123 | 161 | 382.79 | 0.1887 | 32.5 |

The average grain size $\langle D_{hkl} \rangle$, which is averaged over the Debye-Scherrer rings, at a given instant during the deposition is estimated using Debye-Scherrer method [24] and is given by:

$$\langle D_{hkl} \rangle = \frac{1}{N(\Delta\phi_i)} \sum_{\Delta\phi_i=1}^{N} \frac{0 \cdot 9 \times \lambda}{\beta(\Delta\phi_i) \times cos\theta(\Delta\phi_i)} \quad (1)$$

$N(\Delta\phi_i)$ is the number of azimuthal bins. $\beta(\Delta\phi_i)$ is the full width at half maximum of the pattern of individual intensity vs. 2θ plots for a given value of $\Delta\phi_i$ and θ is the corresponding half of the scattering angle (2θ). λ is the wavelength of the x-ray beam and is 0.168 Å for this study.

The ex-situ characterization of the films was carried out by using scanning electron microscopy (SEM, Leo 1550 Gemini with an electron acceleration voltage of 4-5 kV), transmission (TEM) and scanning transmission (STEM) electron microscopy (both performed on a FEI Tecnai G2 microscope operated at 200 kV). STEM was combined with both high angle annular dark field (HAADF) and energy dispersive x-ray spectroscopy (EDS) detectors. Prior to analysis by (S)TEM suitable thin specimens were prepared by using a cross beam focussed ion beam (FIB) instrument (Zeiss EsB 1540) using the standard lift out approach[25]. Mechanical properties of the films such as hardness, and elastic recovery was obtained with a nanoindenter equipped with Berkovich diamond tip. The maximum load of the nanoindentations was set to 10 mN so the indentation depth is less than 10% of the total film thicknesses. The hardness values were quantified by the method proposed by Oliver and Pharr [26]. At least 30 indents were used to calculate the mean values of hardness and their standard deviation.

### III. Results and Discussion

Fractured cross-section SEM micrographs of the films are presented in Figure 1. A typical nano-columnar thin film morphology is seen for samples deposited below 750 °C whereas the film deposited at $T_s$ = 750 °C has a nanocrystalline like morphology (Figure 1(d)). The $Al_{0.50}Ti_{0.50}N$ thin film deposited at $T_s$ = 250 °C shows a variation in film thickness across the film (Figure 1 (a)) indicating a cohesive failure of the film during its deposition. A plan view SEM inspection of the films confirms the delamination of the film from the substrate. From the visual inspection of the SEM images, the width of the nanocolumns is maximum for film deposited at $T_s$ = 450 °C, and minimum for film deposited at $T_s$ = 750 °C.

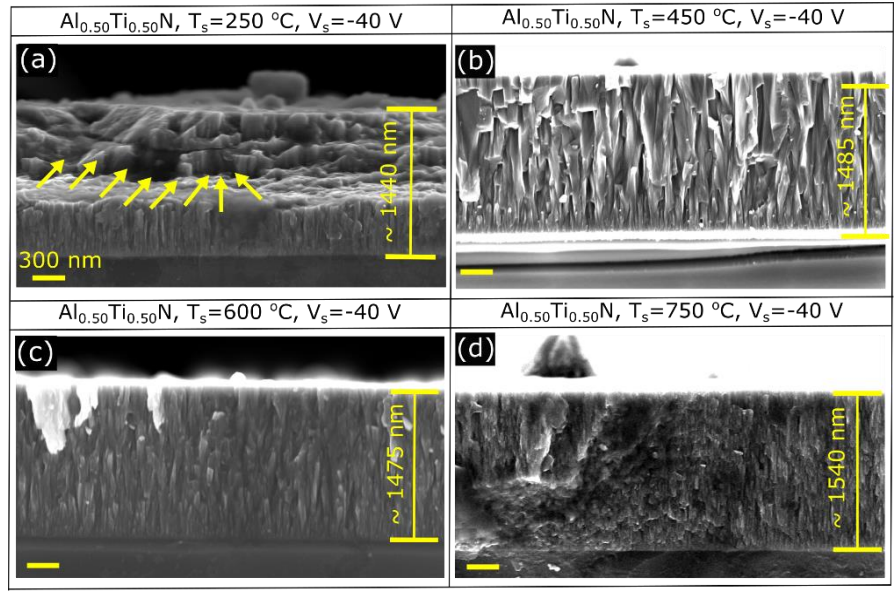

Figure 1 shows the f*ractured cross-sectional FESEM micrographs of Al$_{0.50}$Ti$_{0.50}$N films deposited at substrate temperatures of 250 °C (a), 450 °C (b), 600 °C (c), and 750 °C (d). The yellow-coloured arrow-marked region in figure (a) shows a cohesive failure region in the film.*

The 2D x-ray diffractograms reveal that for all the $T_s$ that are discussed here, the Al$_{0.50}$Ti$_{0.50}$N films crystallized in an FCC structure (see Figure 2). It can be clearly seen from Figure 2 that in the films deposited with $T_s$ = 250 °C, there is a preferred c-200 orientation in the growth direction (GD), while the c-111 orientation is preferred at φ≈35° (see Figure 2 (a)). With increase in $T_s$ to 450 °C and 600 °C, the diffraction signal from c-111 is more randomly distributed in φ. The diffraction pattern recorded for the Al$_{0.50}$Ti$_{0.50}$N film deposited at $T_s$=750 °C differs from the others. Firstly, the diffraction intensity recorded from the sample is relatively low despite using the same geometry and x-ray flux. The diffraction pattern shows that both c-111 and c-200 are weakly textured. However, the most intense diffracted intensity is seen at φ ≈ 54° for c-111 and φ ≈ 47° for c-200.

TEM micrographs of the films deposited at $T_s$ = 450 °C and 750 °C are shown in Figure 3. Figure 3 (a1) shows a bright field TEM micrograph of the film deposited at $T_s$ = 450 °C. The micrograph reveals 50-100 nm wide nanocolumns with no evidence of voids or under-dense grain boundaries. For the first ≈ 300 nm of deposition, a larger number of narrower columns are seen, consistent with our previous observations [21] and a result of competitive grain growth. SAED patterns were recorded at four different locations along the GD and marked as coloured circles in Figure 3 (a1). The SAED pattern recorded at the bottom of the film (green circle, Figure 3 a6) show a preferential c-200 orientation along the GD. There is however an azimuthal spread of the c-200 diffraction signal, indicating small grains with a mosaicity along the GD. The SAED pattern recorded at the top of the film (see orange coloured circle in Figure 3 (a1) and Figure 3 (a3)) shows a spotty diffraction pattern, indicating larger grain sizes with film growth. To verify that the diffraction pattern is representative of the film structure, we recorded SAED patterns from multiple locations on the top part of the film and found that all patterns were consistent. The intense diffraction spots corresponding to c-200 are slightly tilted away from the GD, agreeing well with our in-situ XRD data (see Figure 2 (b)). In this case, a combination of 111 and 113 crystal planes from FCC-Al$_{0.50}$Ti$_{0.50}$N appear to be preferred along the GD. Similar observations are made from the SAED pattern recorded from the middle of the films (see Figure 3 (a4)).

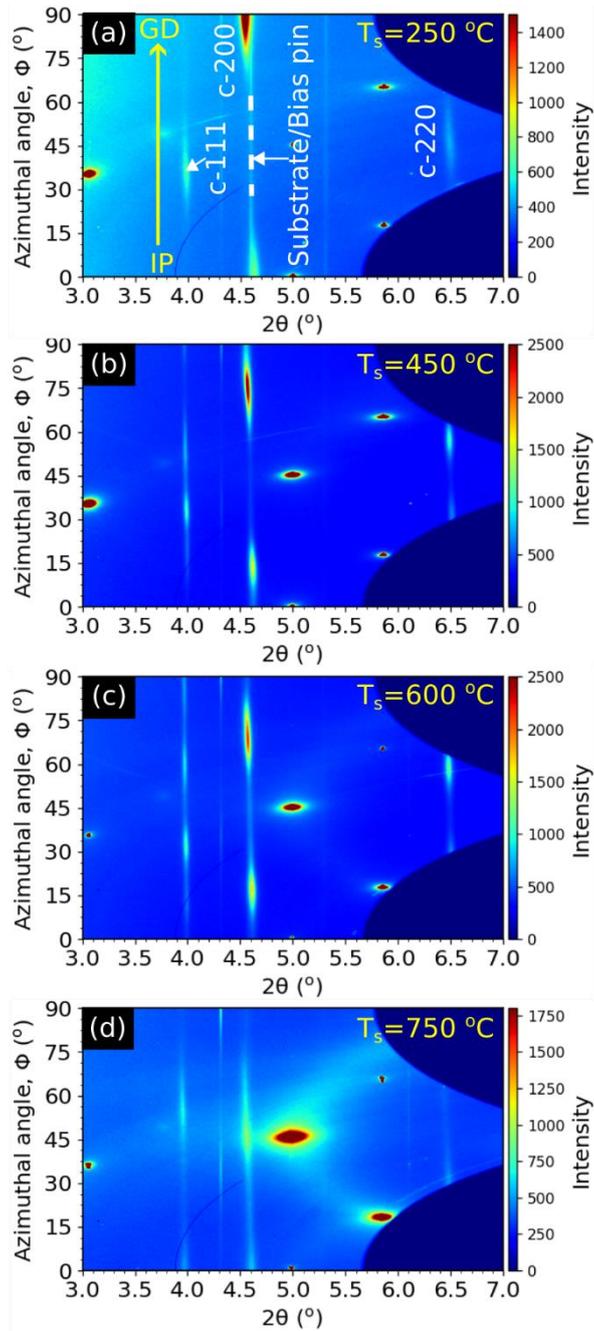

Figure 2 *shows the polar transformed 2D x-ray diffractograms of $Al_{0.50}Ti_{0.50}N$ films deposited at a substrate temperature of 250 °C (a), 450 °C (b), 600 °C (c), and 750 °C (d) at the end of their depositions. For φ = 90° and 0°, the growth direction (GD) and in-plane (IP) directions are assigned.*

Figure 3 (b1) represents bright field TEM images of the film deposited with $T_s$ = 750 °C and reveal a densely packed nanocrystalline-like morphology. The SAED patterns recorded at different film thickness (see coloured circle in Figure 3 (b1)) are presented as Figure 3 (b3-b6). At the top of the film, the pattern consists of defined arcs rather than discrete spots as for the $T_s$ = 450 °C film. This indicates that the film is composed of smaller. Irrespective of the film thickness, the c-220 crystal planes of FCC-(Al, Ti)N are predominantly textured along the GD, while c-111, and c-200 crystal planes textured close to 35-45° tilted away from the GD (see Figure 3 (b3-b6)), in agreement with our in-situ XRD findings (see Figure 2 (d)). It was also observed that diffractograms from both c-111 and c-200 planes appeared very close to each other in azimuthal angle, φ. This suggests that there was some coherency between both c-111 and c-200 oriented grains in the film. Besides the diffraction pattern from FCC-(Al, Ti)N, we observed a faint diffraction signal from 0002 crystal planes (h-0002) of hexagonal-closed-packed (HCP)-(Al, Ti)N (see the red coloured arrow marked in Figure 3 (b3-b6)). We also note that the diffraction from c-111 and h-0002 have the same azimuthal angle (Figure 3 (b3-b6)), suggesting that the c-111 and h-0002 crystal planes share the same orientation. STEM-EDS mapping of Si, Al, Ti, and N, shown as Figure 3 (a2) and (b2) respectively, did not reveal any measurable compositional modulation in the films, and Al, Ti, and N form a uniform solid solution over the film thickness.

Figure 4 shows the texture evolution of c-200 planes of (Al, Ti)N thin films deposited at different $T_s$. The film deposited at $T_s$ = 250 °C shows that c-200 crystal planes are aligned along the GD (φ=90°) throughout the film thickness, $t_f$ (see Figure *4* (a)). Contrary to this, as mentioned previously, film deposited at $T_s$ = 450 °C and 600 °C show that c-200 crystal planes are textured at a certain tilt angle to the GD. Interestingly, as the $t_f$ increases, the tilt angle of the c-200 texture also increases. The change in tilt angle as the $t_f$ increases from 300 nm to ~1450 nm is estimated to 7° and 10° for film deposited with $T_s$ = 450 and 600 °C respectively (see (b-c)). The results agree well with the SAED pattern recorded from thin film deposited with $T_s$ = 450°C (see Figure *4* (a3-a6)). The c-200 of the film that was deposited at $T_s$ = 750 °C have a texture that is relatively broad along φ, but the intense spots only occur at φ ≈ 45° and stay consistent throughout the thickness of the film (see Figure 2 (d)). Thus, two important observations are made from *ex-situ* data:

(i) In polycrystalline thin films of (Al,Ti)N, if c-200 crystal planes are tilted away from the GD, a mixed of c-111+c-113 textures develops along the GD.
(ii) There is a consistent increase in the tilt angle of c-200 planes of grains that are formed far from the interface of the films compared to towards the interface of the films.

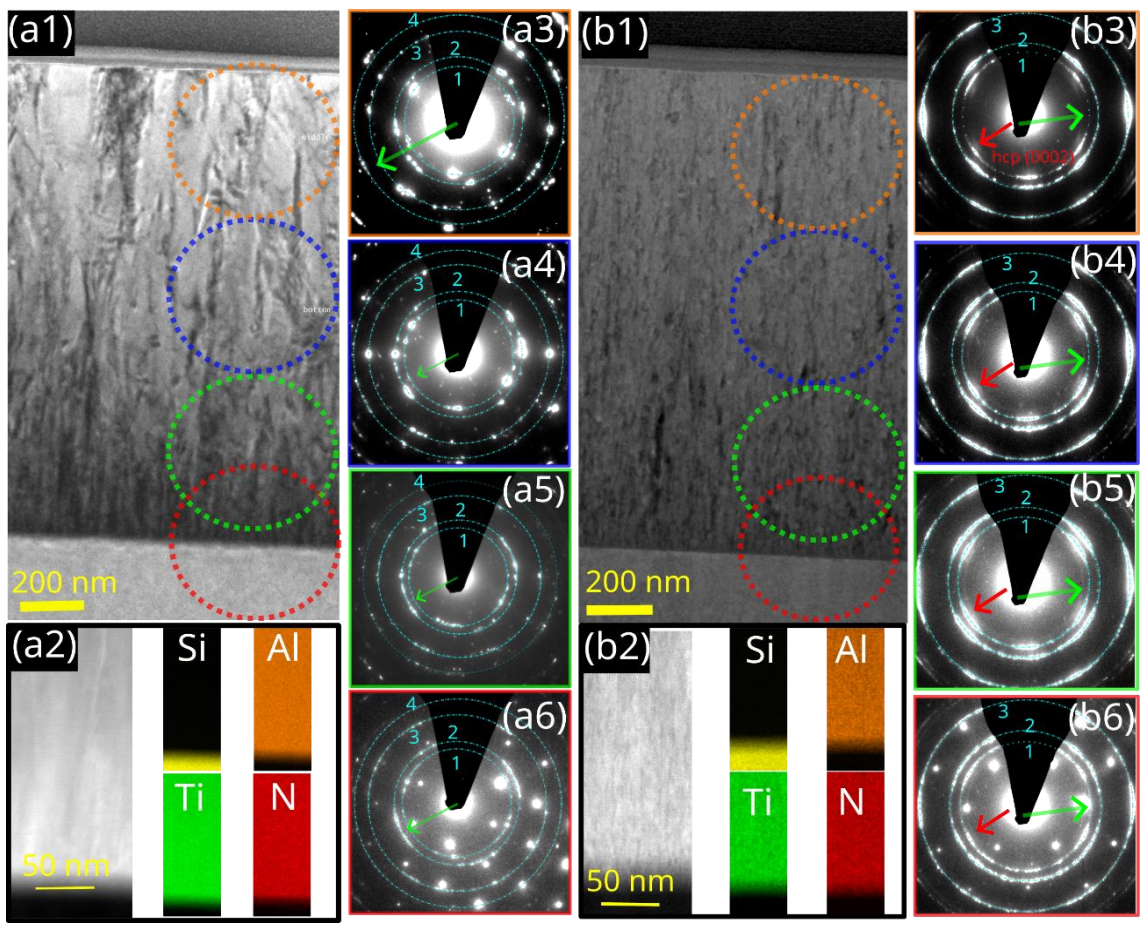

Figure 3 *shows TEM micrographs of the $Al_{0.50}Ti_{0.50}N$ thin films deposited with $T_s$ of 450 °C (a1-a6) and 750 °C (b1-b6). Figures a1, b1 are BF-TEM micrographs; a2, b2 are STEM micrographs and corresponding EDS maps, and a3-a6, b3-b6 are the SAED patterns recorded at various locations (shown as dotted-coloured circles at Figure a1, b1). The diffraction signal present on the circumference of circles marked as 1, 2, 3, and 4 on Figures a3-a6 and b3-b6 are identified as 111, 200, 220, and 113 crystal planes of FCC-(Al, Ti)N. The green arrows in a3-a6 and b3-b6 represent the growth direction. The red arrows in Figures b3-b6 show diffraction from the 0002 planes of HCP-(Al,Ti)N.*

The texture formation of c-200 crystal planes along the GD is well documented in the literature, and is often attributed to the minimization of surface energy, strain energy, and the channelling effect of ions at lower substrate bias voltages (e.g., -40 V) that are observed in films deposited at $T_s$ = 250 °C. However, the microscopic origin of the off-axis tilt of 200 crystal planes away from the GD has been debated throughout the literature. Rafaja et al. [27] attribute the inclined preferred orientation in AlTiN deposited by arc evaporation to the geometry of the deposition system with a non-rotating substrate. Wahlström et al.[28] argued the tilted texture of magnetron sputtered AlTiN films is due to the oblique path between the substrate and magnetron targets. Je et al. [29] attributed the ±5° tilt of the 002 planes in TiN to competitive growth of 111 and 200 oriented grains, and to maintain the overall growth direction of 111 oriented grains close to the surface normal, the 200 oriented grains are tilted away. The inclined texture of nanostructured $Al_{1-x}Ti_xN$ is attributed to twin faults developed in the lattice by Kalss et al. [30]. Falub et al. argued an interdependence of texture and stress in arc evaporated AlTiN thin films and suggested stress can influence the texture formation, and the tilt angle increases with an increase in the level of stress in the films [17]. A similar interpretation is made for CrN [31–33], and $Ti_{0.67}Al_{0.33}N$ [34] thin films, where the transition of texture from c-200 to c-113 is attributed to a decrease in stress and film density. Karimi et al. suggest that competitive growth promotes the formation of c-111-oriented grains, and to accommodate growth of this c-111-oriented grain, the c-200-oriented grains are tilted away from the GD [35]. It was observed that with an increase in substrate temperature, the tilt angle of c-200 reduces owing to an increase in adatom mobility at higher substrate temperatures [35].

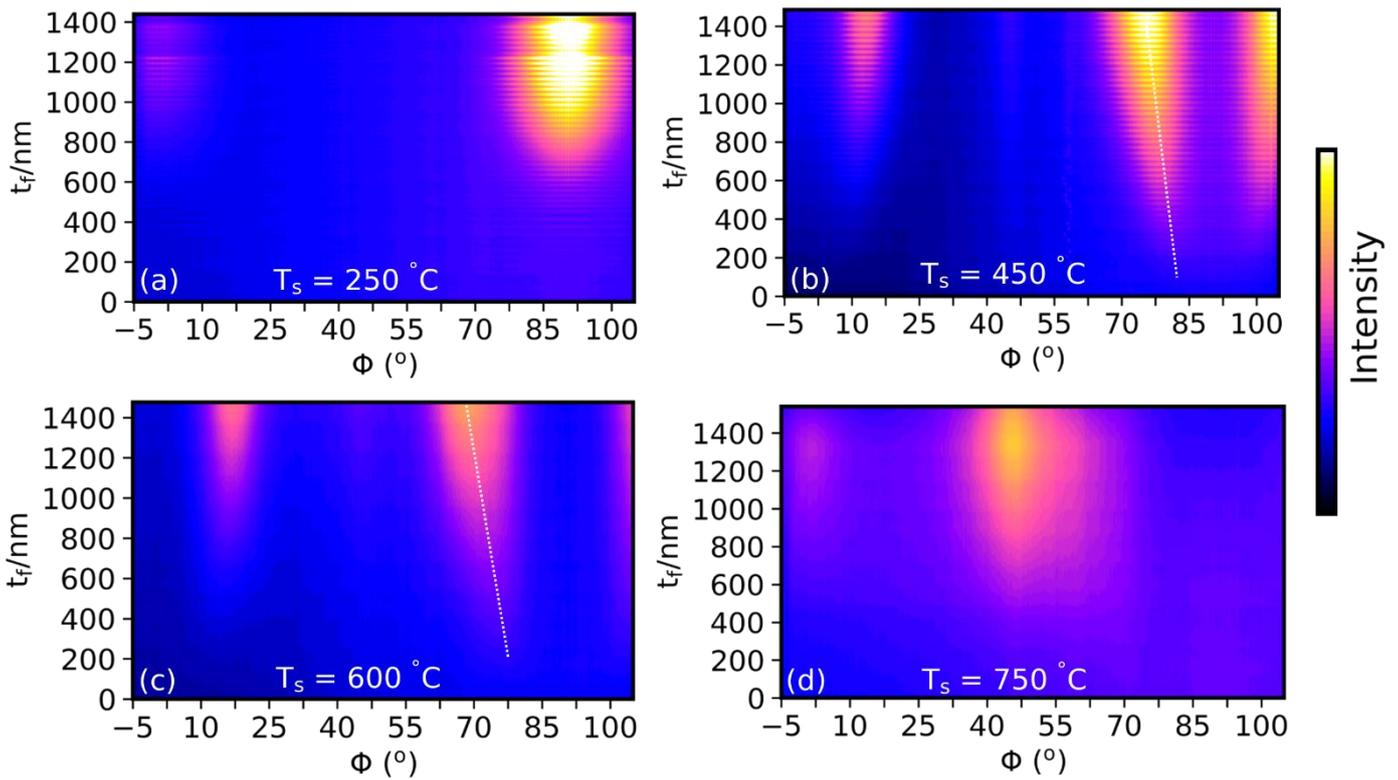

Figure 4 shows the evolution of texture of 200 crystal planes of FCC-(Al,Ti)N along the azimuthal angle(φ) for film deposited with a substrate temperature, $T_s$, of 250 °C (a), 450 °C (b), 600 °C (c), and 750 °C (d). The dotted lines in (b) and (c) are guided lines for change in Φ as $t_f$ increases.

In our study, the off-axis tilt of c-200 texture cannot be attributed to the oblique incidence of metal vapours toward the substrate, as previously argued [27,28,36], because perfect alignment of c-200 texture along the GD is observed with film deposited at the same geometrical configuration (i.e., 35° inclination angle), but at a relatively lower $T_s$ of 250 °C. Our in-situ stress analysis (see Figure 5 (a-d)) revealed that the stress level in films is also not responsible for the tilt of the c-200 texture as proposed by Falub et al. [17] and Karimi et al. [34]. The $Al_{0.50}Ti_{0.50}N$ thin film deposited at $T_s$ = 250 °C shows high stress state (see Figure 5 (a)), but the c-200 planes are perfectly textured along the GD. Between films deposited at $T_s$ = 450 °C and 600 °C, the latter shows a lower stress (Figure 5 (b-c)) and a higher off-axis tilt angle of c-200 than the former ones, contradicting the predictions by Falub et al.[17] and Karimi et al. [34]. A consistent increase in tilt angle with an increase in substrate temperature is also in contradiction to the model proposed by Karimi et al [34]. We propose that in these samples, along with the competitive growth mechanisms, other factors such as that which causes film surface roughening (e.g., the shadowing effect) during deposition [37] are together responsible for the off-axis tilt of c-200 in films deposited with $T_s$ = 450 °C and 600 °C.

To the best of our knowledge, there is no report on the dynamics of texture evolution that are determined in real time during the film depositions. The literature has never documented a constant rise in off-axis tilt with film thickness except observations made by Kubec et al. (Keckes et al., 2018)[37] in a multi-layered TiN and $SiO_x$ thin film and was attributed to the difference in nucleation nature between the substrate and $SiO_x$ buffer layer and between two TiN layers. At this point, our hypothesis is that the consistency of the off-axis tilt of the 002 texture of $Ti_{0.50}Al_{0.50}N$ thin films deposited at $T_s$ of 450 °C and 600 °C is controlled by the rise in surface roughness(Shetty & Karimi, 2011)[34] and/or formation of twinned crystallites with the film thickness.

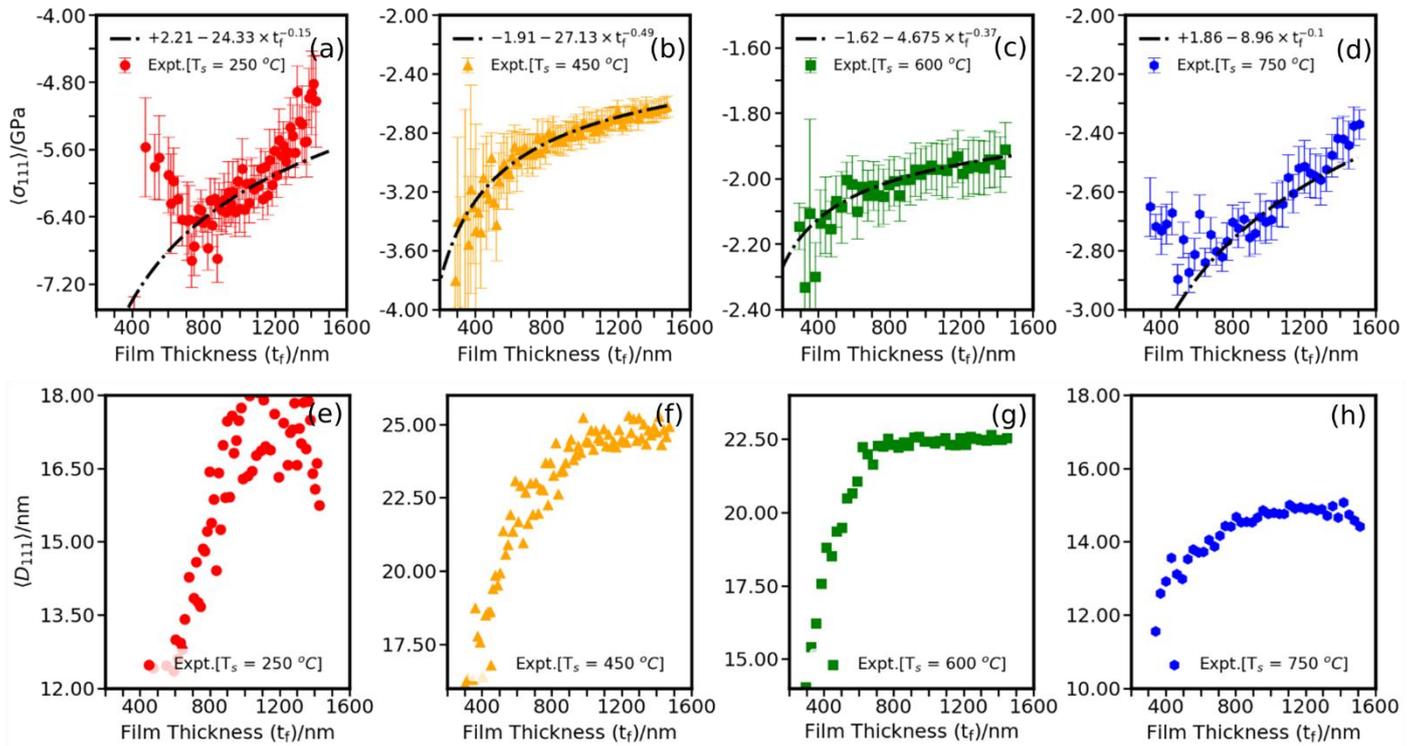

Figure 5 (a-d) shows the evolution of biaxial stress $\langle\sigma_{111}\rangle$ as a function of film thickness, $t_f$, of $Al_{0.50}Ti_{0.50}N$ at a substrate temperature of 250 °C, 450°C, 600°C and 750 °C respectively. Figure (e-h) shows the temporal evolution of the average crystallite size $\langle D_{111}\rangle$ of the films deposited at a substrate temperature of 250 °C, 450 °C, 600 °C, and 750 °C, respectively.

The formation of c-220 texture, as seen in the film deposited with a $T_s$ of 750 °C, in TMNs' thin films deposited with PVD and CVD is seen previously. Often in PVD deposition process, the c-220 textured was seen, when depositions were carried out with high energetics ions [39–41] as <220> direction has the high degree of channelling effect in FCC lattice. While in CVD-deposited thin films, the formation of c-220 texture is attributed to the formation of twinned crystallites at temperatures lower than 1200 °C [42], and the formation of twinned crystallites is attributed to the lower adatom mobility. To gain a microscopic insight into the c-220 texture formation in $Al_{0.50}Ti_{0.50}N$ film with $T_s$ = 750 °C, we probed the atomic scale crystal using HR-TEM. As the c-220 textured from at the very initial stages of depositions (as seen from the SAED pattern, see Figure 3 (b6)) towards the interface of the film and substrate. Below a film thickness of 5 nm, the lattice planes could not be clearly resolved (see Figure 6(a)). However, beyond that, the formation of specific crystalline grains is seen. The formed grains are found to be faceted (see the yellow dotted lines in Figure 6 (a) and the angle between the normal of the crystal facets and the normal of the film/substrate interface varies from 35° to 45°. The FFT pattern (see Figure 6 (b)) also showed that the strongest intensity of the c-111 lattice planes was tilted away from the GD, along with a lower intensity of both c-111 and c-200 crystal planes being aligned along the GD. This established formation of c-111 crystal facets at very initial stages of depositions ($t_f \approx$ 15 nm) starts. A similar analysis for the film deposited with Ts= 450 °C showed an intense texture of c-200 along the GD. The formation of these crystal facets could be driven by the change in supersaturation of Al in $Al_{0.50}Ti_{0.50}N$ at this high temperature ($T_s$ = 750 °C) [43]. These newly formed facets are responsible for the further nucleation of tilted, 111-oriented grains. And since the angle between c-111 and c-220 crystal planes in the FCC lattice is 35.3°, the c-220 planes are textured along the GD in this case. The frequent re-nucleation induced nanocrystalline surface morphology, relatively higher growth rate, and formation of minute HCP-$Al_xTi_{1-x}N$ in this film suggest the film is grown under high supersaturation [44,45]. The appearance of the diffraction pattern of h-0002 crystal planes of HCP-$Al_xTi_{1-x}N$ and c-111 of FCC-$Al_{0.50}Ti_{0.50}N$ along the same azimuthal angle (as seen in Figure 3 (b3-b6)) is commensurate with the symmetry and low lattice mismatch between these planes [46].

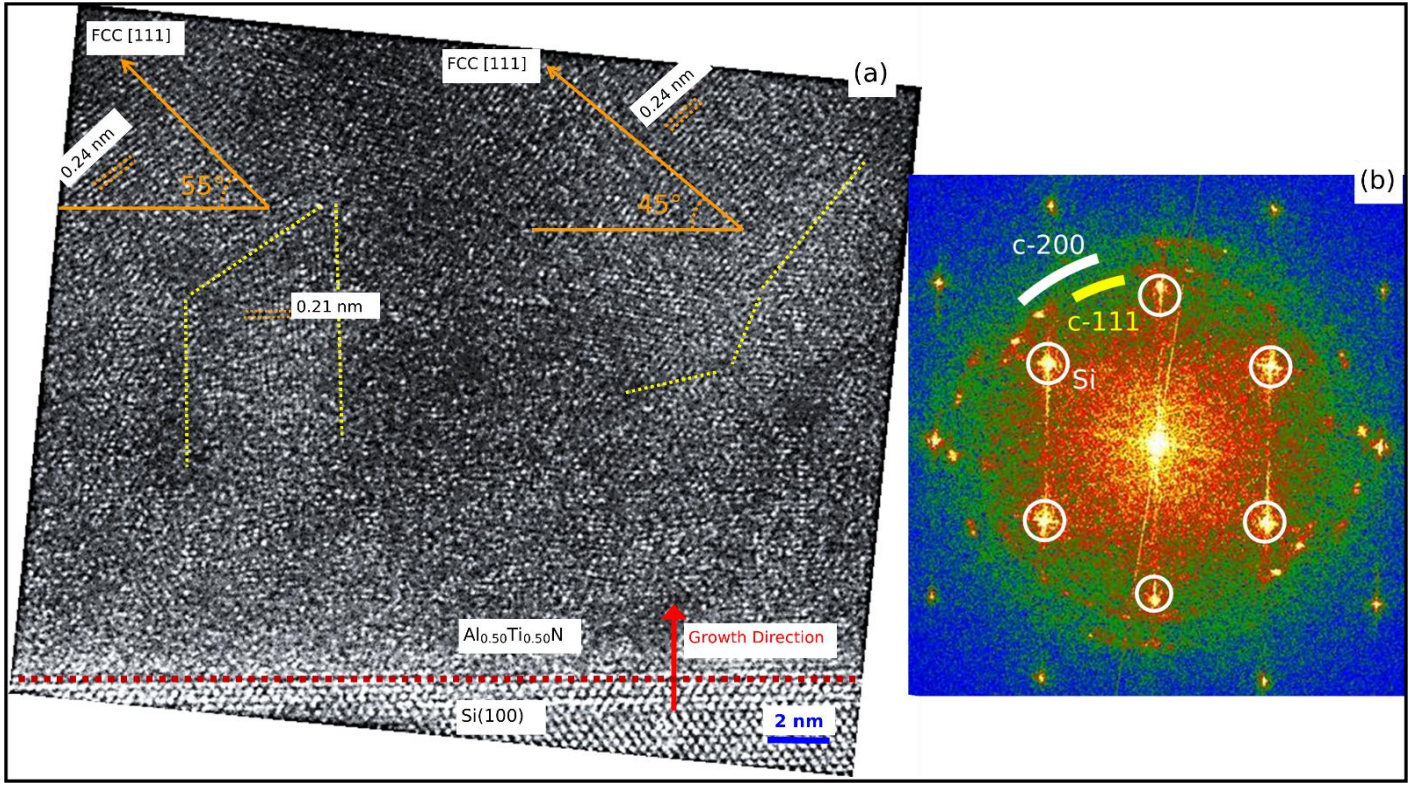

Figure 6 is the HRTEM image (a) and Fast Fourier Transformation (FFT) pattern (b) of the interface region of the $Al_{0.50}Ti_{0.50}N$ thin film grown on Si (100) substrate deposited at $T_s$ = 750 °C.

Figure 5 shows the evolution of biaxial stress ($\langle\sigma_{111}\rangle$) and average grain/domain size ($\langle D_{111}\rangle$) as a function of $t_f$. Because of the low intensity of the diffraction signal, data could not be extracted below film thicknesses of ~300 nm. For the film deposited with $T_s$ = 250 °C, $\langle\sigma_{111}\rangle$ increases consistently until $t_f \approx 800$ nm, but at higher $t_f$ the $\langle\sigma_{111}\rangle$ decreases as $t_f$ increases (see Figure 5 (a)) suggesting a compressive stress relaxation with film thickness. The films deposited at $T_s$ = 450 °C and 600 °C show a clear power-law like relaxation of $\langle\sigma_{111}\rangle$ with $t_f$ (see Figure 5 (b,c)). The power law fitting reveals that the $\sigma_0$, c and γ are -1.91 GPa, -27.13 GPa nm$^\gamma$, and 0.49, respectively for $T_s$ = 450 °C and -1.62 GPa, -4.675 GPa nm$^\gamma$ and 0.37, respectively for $T_s$ = 600 °C. The evolution of $\langle\sigma_{111}\rangle$ with $t_f$ for the film deposited with $T_s$ = 750 °C is different than Ts = 450 and 600 °C the previously discussed ones. Between $t_f \approx 300$ nm and $t_f \approx 750$ nm the $\langle\sigma_{111}\rangle$ increases, while at larger $t_f$ a consistent relaxation of $\langle\sigma_{111}\rangle$ is observed (see Figure 5 (d)).

The evolution of average grain/domain size $\langle D_{hkl}\rangle$ as a function of $t_f$ is shown in Figure 5 (e-h). The film deposited with $T_s$ = 250 °C shows a consistent increase in $\langle D_{111}\rangle$, reaching 18.2 nm at $t_f$ = 1250 nm, beyond which a scattered value of $\langle D_{111}\rangle$ is seen (see Figure 5 (e)). When $T_s$ is raised to 450 °C, the estimated $\langle D_{111}\rangle$ increases with thickness to 25.2 nm at $t_f$ = 1485 nm (see Figure 5 (f)). For the coatings grown at 600 °C, the maximum grain size of 22.2 nm is reached already at $t_f$ = 620 nm, beyond which it remains approximately constant (see Figure 5 (g)). At $T_s$ of 750 °C, the change in $\langle D_{111}\rangle$ is very small, and it increases from 11.5 nm to a maximum of 14.6 nm, as $t_f$ increases from 300 nm to 1540 nm (see Figure 5 (h)).

The generation of compressive stress in polycrystalline thin films is widely reported in the literature and is attributed to densification of grain boundaries and the generation of defects in the bulk by high energy ions [47,48]. The generation of compressive stress in our films agrees with the literature, as the films are deposited with a -40 V of substrate bias. However, the stress relaxation associated with an increase in total film thickness, $t_f$, needs to be explained. Also, the microscopic origin of anomalies in the $\langle\sigma_{111}\rangle$ evolution in films deposited at $T_s$ = 250 °C and 750 °C needs to be examined. The relaxation of compressive stress in films with an increase in film thickness indicates that a tensile component is being generated, often due to grain growth or larger average domain sizes. Since the deposition temperatures are well below the melting point of AlTiN (~2949 °C (Toth, n.d.)[48]) it is unlikely the grain growth due to bulk diffusion. Thus, we suggest the larger average grain sizes with increase in $t_f$ originates from the kinetics effect (see Figure 5). The contribution from the tensile component of grain growth or increase in average grain size and the compressive component due to grain boundaries and adatom attachment to 2D islands formed at the surfaces of grains, were deconvoluted by Thompson et al. *(Yu & Thompson, 2014)[49]*. The instantaneous tensile stress due to grain growth ($\sigma_{in}^{gg}$) is given by;

$$\sigma_{in}^{gg}(d) = \frac{E}{1-\nu}\frac{\Delta a}{d}\left(\frac{1}{1+\left(\frac{E}{1-\nu}\right)\frac{\Delta a}{d}\frac{1}{\sigma_y}}\right) \quad (2)$$

where $E$, and $\nu$ are Young's modulus and Poisson's ratio, respectively. The $d$, $\Delta a$, and $\sigma_y$ represent the average grain size, excess free volume associated with grain boundaries, and biaxial tensile stress yield limit respectively. The authors [51] further argued that the instantaneous stress generated by the newly deposited surface layer ($\sigma^*(t)$) is the ratio of the capture probability of adatoms to the grain boundary ($N_{adatoms-GB}$) and to the newly formed 2D islands ($N_{adatoms-2Disland}$). Thus, the compressive component of instantaneous stress is;

$$\sigma_{in}^{comp} = -\frac{E}{1-\nu}\frac{2p\delta}{d} \quad (3)$$

where $p$ is the capture probability of the adatoms by the grain boundary, and $\delta$ is the average distance between the 2D islands and the grain boundaries. In the smaller grain limit of $d < L_{island}$, $L_{island}$ being the 2D island spacing, the $\sigma_{in}^{comp}$ is independent of grain size, $d$. The transition of $\sigma_{in}^{comp}$ being independent of $d$ to its inverse dependence to $d$ occurs at $d = L_{island}$. We applied the model to the evolution of $\langle\sigma_{111}\rangle$ with increasing film thickness $t_f$. The instantaneous compressive stress was calculated through $\sigma_{in}^{comp} = \langle\sigma_{111}\rangle - \sigma_{in}^{gg}(d)$ for all four films. Even though the growth temperatures used for this study were relatively low for bulk diffusion that results into grain growth, for completeness of the model we estimate $\sigma_{in}^{gg}(d)$ in following way. $\sigma_{in}^{gg}(d)$ (mentioned as $\sigma_{111}^{gg}$ in Figure 7) was calculated using $\Delta a = 1$ Å, and $\sigma_y = 830$ MPa (the maximum converged tensile stress recorded from a similar study [21]) in Eqn. 2. The instantaneous $d$ was approximated to $\langle D_{111}\rangle$. Thus, the estimated $\sigma_{in}^{gg}(d)$ represent the maximum tensile stress generated by grain growth. For $E$ and $\nu$, the high temperature values of $E_{111}$ and $\nu_{111}$ were used.

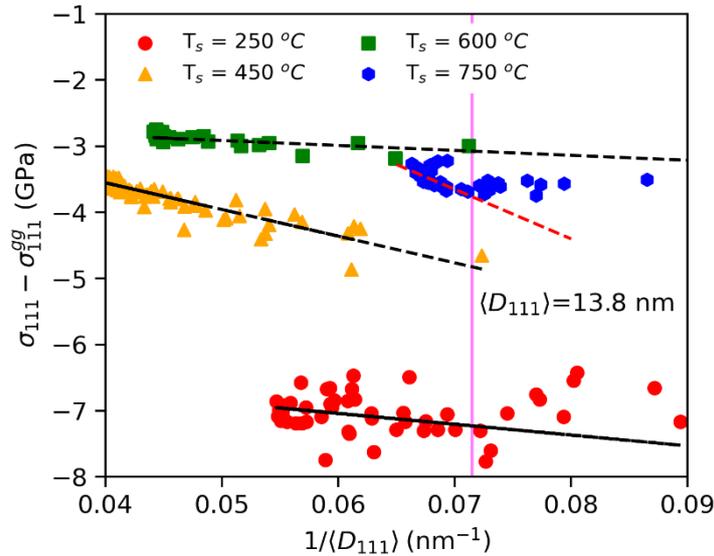

Figure 7 shows instantaneous compressive stress ($\sigma_{111} - \sigma_{in}^{gg}(d)$ as a function of inverse of grain size ($\langle D_{111}\rangle$). The dashed/solid lines over experimental scatter data points shows linear fitting of the experimental $\sigma_{111} - \sigma_{in}^{gg}(d)$ vs inverse of $\langle D_{111}\rangle$.

The estimated $\sigma_{in}^{comp}$ vs $\frac{1}{d}$ (or $\frac{1}{\langle D_{111}\rangle}$) is shown in Figure 7. At $\frac{1}{\langle D_{111}\rangle} = 0.072$ nm$^{-1}$ the behaviour changes. At values lower than 0.072 nm$^{-1}$, there is a linear dependence of $\sigma_{in}^{comp}$ on $\frac{1}{\langle D_{111}\rangle}$, whereas for larger values, $\sigma_{in}^{comp}$ is independent of $\frac{1}{\langle D_{111}\rangle}$. Thompson et al. [50] proposed that the transition from inverse dependence to independence of $\sigma_{in}^{comp}$ occurs at $d = L_{island}$. Thus, we identify the $L_{island}$ to 1/0.072 nm$^{-1}$=13.9 nm. We also estimated $\sigma_{in}^{gg}(d)$ by varying $\Delta a$ from 0.5 Å to 1 Å and $\sigma_y$ from 500 MPa to 1.0 GPa and analysed the depencence of $\sigma_{in}^{comp}$ to $\frac{1}{\langle D_{111}\rangle}$. The obtained results are similar as discussed above. As a result, we suggest that, for grain sizes smaller than 13.9 nm, the compressive stress, $\langle\sigma_{111}\rangle$, does not relax with increasing in film thickness, despite an increase in grain size with film thickness. In this grain size regime, the capture probability of adatoms at the grain boundary is higher than that of the 2D islands formed on top of grains, leading to the generation of compressive stress. With grain size larger than 13.8 nm, the effect is reversed, and the capture probability of adatoms on 2D islands increases as grain size increases, leading to a reduction in the magnitude of compressive stress and hence a small stress relaxation observed in our films. The calculated $L_{island}$ is in close agreement with our earlier predictions of the mean grain size ($D_0 = 13.4 \pm 3.6$ nm) at which discrete islands shift to continuous films in $Al_xTi_{1-x}N$ coatings [21]. The computed $L_{island}$ and $D_0$ are within the range of 2D island spacing anticipated by classical nucleation theory [52].

The relatively smaller grain size in film deposited at $T_s$ = 250 and 750 °C compared to $T_s$ = 450 and 600 °C can be understood as follows:

> (i) Because of the lower substrate temperature, $T_s$, the adatom mobility as well as the grain boundary mobility are smaller, and hence the rate of increment of average grain size due to growth kinetics is lower. This leads to a smaller average grain size and, consequently, relatively narrow nanocolumns and an intercolumnar gap (as seen in Figure *1* (a)).
> 
> (ii) Adatom mobility and grain boundary mobility both rise as the substrate temperature rises to 450 and 600 °C, causing increase in average grain sizes as film thickness increases and the emergence of wider and denser nanocolumns (as seen in Figure *1* (b-c) and Figure *3*(a1)).
> 
> (iii) At $T_s$ = 750 °C, the development of smaller grain size is a little trickier. The nanocrystalline microstructure shows that during thin film growth, grain nucleation occurs frequently. It is possible that the formation of hexagonal $Al_xTi_{1-x}N$ at grain boundaries driven by thermodynamics, as previously observed, is what limits kinetics induced grain growth, as suggested by the occurrence of a faint diffraction from HCP-$Al_xTi_{1-x}N$ 0002 planes.

In $Al_{0.50}Ti_{0.50}N$ thin films deposited at $T_s$ = 250 °C, the evolution of $\langle\sigma_{111}\rangle$ deviates from power-law after a total film thickness of 1250 nm, and the evolution of $\langle D_{111}\rangle$ becomes scattered (see Figure 5 (a) and (e)). We also observe a discontinuity at the same $t_f$ in the evolution of the integrated intensity of diffraction patterns from crystal planes c-111 and c-200. The SEM also demonstrates the film's lack of coherence (see Figure 1 (a)). Our earlier work made a similar set of observations, which we attribute to cohesive failure of films due to high stress fields. The critical stress ($\sigma_{cr.}$) and critical film thickness ($h_{cr.}$) values, which correspond to the point(s) of discontinuity in the graph of $\langle\sigma_{111}\rangle$ vs $t_f$, can be used to calculate the plain strain fracture toughness ($K_{IC}$) [21,53,54]. The $K_{IC}$ is estimated to be $4.45 \pm 0.4$ MPa.m$^{1/2}$ for an approximate $\sigma_{cr.} = \sigma_{111}$ (at $t_f$ = 1250 nm) and $h_{cr.}$= 1250 nm. This estimate is very similar to the $Al_{0.50}Ti_{0.50}N$ films deposited at a relatively higher bias voltage (-80 V), but it is slightly higher than previously reported values for $Al_{0.60}Ti_{0.40}N$ [1] and $Al_{0.46}Ti_{0.54}$ [55].

To characterize the effect of different microstructures to the mechanical properties of $Al_{0.50}Ti_{0.50}N$ films, we measured the mechanical hardness and elastic recovery from thin films deposited with $T_s$ = 450, 600, and 750 °C. The recorded hardness values are $30.4 \pm 1.4$, $32.7 \pm 0.9$, and $30.9 \pm 1.4$ GPa. The elastic recovery is measured with a load of 10 mN and is found to be 43.7%, 47.6%, and 43.8% for film deposited at $T_s$ = 450, 600, and 750 °C, respectively. As the uncertainty in the estimation of hardness is larger than the difference between samples, a definitive conclusion on the effect of microstructures on the mechanical hardness of $Al_{0.50}Ti_{0.50}N$ thin film deposited with cathodic arc evaporation could not be drawn conclusively.

## IV. Summary

We used in-situ 2D synchrotron x-ray diffraction to investigate the impact of substrate temperature on the in-situ internal stress, crystallographic texture, and grain/crystallite sizes of $Al_{0.50}Ti_{0.50}N$ thin films deposited on 100 surface of silicon using the reactive cathodic arc evaporation method. The thin films exhibited a single-phase FCC crystal structure and a nanocolumnar morphology, when they were formed at a substrate temperature of less than 750 °C. Small amounts of HCP phased $Al_xTi_{1-x}N$ formed at substrate temperature of 750 °C, along with the dominating FCC phase, resulting in a nanocrystalline-like morphology. We discovered that the temperature of the substrate has a significant influence on the development of crystallographic textures. The 200 crystal planes are oriented parallel to the film-substrate interface at low substrate temperature (250 °C). The 200 crystal planes are tilted away from the direction perpendicular to the normal of the film/substrate interface as the substrate temperature rises to 450–600 °C, and a combination of 111 and 113 crystallographic texture is generated along the growth direction. When the substrate temperature reaches 750 °C, a 220 crystallographic texture begins to form. We also see that the off-axis tilt of 200 crystal planes increases between 7°-10° as the film thickness increases from ~250 nm to ~1400 nm. In conjunction with an increase in average grain size with an increase in overall film thickness, we observe a power-law-like behaviour of biaxial stress relaxation with increasing film thickness for films grown at 450–600 °C. Our investigation showed that the biaxial compressive stress is independent of grain size up to a certain size (determined as ~ 14 nm for $Al_{0.50}Ti_{0.50}N$), after which the stress level scales with the inverse of average grain sizes. The highest degree of biaxial compressive stress, which leads to film cohesive failure, is seen in the thin film of Al$_{0.50}$Ti$_{0.50}$N formed at a substrate temperature of 250 °C. The plain strain fracture toughness of Al$_{0.50}$Ti$_{0.50}$N is estimated to be 4.45 ± 0.4 MPa.m$^{1/2}$, which is in good agreement with our earlier estimates.

**Acknowledgments**

The synchrotron experiments were conducted at DESY under the proposal I-20210060EC. We acknowledge the financial support from the Swedish Research Council (grant no 2017-06701) via the Röntgen Ångström Cluster (RÅC) Frame Program, the competence center FunMat-II supported by Vinnova (grant no 2022-03071), and the Swedish government strategic research area grant AFM (SFO Mat LiU, grant no 2009-00971)